# Resistive Graphene Humidity Sensors with Rapid and Direct Electrical Readout

[1]Anderson D. Smith, [2]Karim Elgammal, [3,*]Frank Niklaus, [2,4,5]Anna Delin,
[3]Andreas C. Fischer, [1]Sam Vaziri, [3]Fredrik Forsberg, [2,6]Mikael Råsander,
[2]Håkan Hugosson, [2,5]Lars Bergqvist, [3]Stephan Schröder, [7]Satender Kataria,
[1]Mikael Östling, [1,7,*]Max C. Lemme

[1]KTH Royal Institute of Technology, Department of EKT, School of Information and Communication Technology, Electrum 229, SE-16440 Kista, Sweden

[2]KTH Royal Institute of Technology, Department of Materials and Nano Physics, School of Information and Communication Technology, Electrum 229, SE-16440 Kista, Sweden

[3]KTH Royal Institute of Technology, Department of Micro and Nano Systems, School of Information and Communication Technology, SE-10044 Stockholm, Sweden

[4]Uppsala University, Department of Physics and Astronomy, Materials Theory Division, Uppsala University, Box 516, SE-75120 Uppsala, Sweden

[5]SeRC (Swedish e-Science Research Center), KTH Royal Institute of Technology, SE-10044 Stockholm, Sweden

[6]Department of Materials, Imperial College London, SW7 2AZ, London, United Kingdom.

[7]University of Siegen, Hölderlinstr. 3, 57076 Siegen, Germany

*Corresponding authors: frank.niklaus@kth.se; max.lemme@uni-siegen.de

**Abstract**

We demonstrate humidity sensing using a change of electrical resistance of a single-layer chemical vapor deposited (CVD) graphene that is placed on top of a $SiO_2$ layer on a Si wafer. To investigate the selectivity of the sensor towards the most common constituents in air, its signal response was characterized individually for water vapor ($H_2O$), nitrogen ($N_2$), oxygen ($O_2$), and argon (Ar). In order to assess the humidity sensing effect for a range from 1% relative humidity (RH) to 96% RH, devices were characterized both in a vacuum chamber and in a humidity chamber at atmospheric pressure. The measured response and recovery times of the graphene humidity sensors are on the order of several hundred milliseconds. Density functional theory simulations are employed to further investigate the sensitivity of the graphene devices towards water vapor. Results from the interaction between the electrostatic dipole moment of the water and the impurity bands in the $SiO_2$ substrate, which in turn leads to electrostatic doping of the graphene layer. The proposed graphene sensor provides rapid response direct electrical read out and is compatible with back end of the line (BEOL) integration on top of CMOS-based integrated circuits.

Solid-state gas sensors have become popular due to their low cost and scalability and have already found their way into a number of different applications.[1,2] Current research focuses on sensors based on metal oxides,[3] semiconductor nanowires,[4–7] carbon nanotubes[4,5] and most recently, solid-state gas sensors based on graphene and graphene oxide (GO).[8–19] One of the notable properties of graphene is its high electrical conductivity, which can be ascribed to the p-orbital electrons, which form π-bonds with neighboring atoms. These π-bonds with their de-localized electrons define the electronic band structure[8] with its high carrier mobility.[20,21] However, the delocalized π-electrons are also sensitive to modifications of their immediate environment. As a consequence, graphene has been shown to be sensitive to a number of different gasses.[22–26] It has also provided the ultimate level of sensitivity by detection of single gas molecules.[27] Several reports have previously studied the influence of humidity on graphene-based devices, but the results reported thus far are limited in range and / or response times.[12,16,17,28,29] Here, we present rapid response resistive humidity sensing using CVD graphene placed on an $SiO_2$ layer for potential solid state sensor applications. The sensors are operational in atmospheric conditions with negligible cross-sensitivity from competing gasses. The humidity sensing mechanism is explained by interactions of the polar $H_2O$ molecules with substrate ($SiO_2$) defects through density functional theory simulations. Chemical vapor deposited graphene was used and the sensor design allows easy integration with CMOS-based circuits, thereby offering a low-cost and highly scalable alternative to conventional humidity sensors for system-on-chip (SoC) solutions.

The process flow for fabricating the graphene sensors is shown in Figure 1a, and a detailed explanation of the fabrication is provided in the methods section. The sensors are fabricated on silicon substrates with thermally grown, 300 nm thick $SiO_2$ layers (Figure 1a-1). First, gold contacts are embedded in the oxide (Figure 1a-2) that can be are used for four-point resistance measurements of the graphene patch (although other metals may be used). Chemical vapor deposited graphene[30] is then transferred from copper foils onto the substrate (Figure 1a-3) and etched into the desired shape (Figure 1a-4). The effective dimensions of the graphene patches are 44 µm in length (the distance between the inner Au contacts) and 80 µm in width. Raman spectroscopy shows the typical graphene G and 2D peaks (Figure 1b)[31] and confirms the successful graphene transfer as well as the absence of a discernable defect peak – demonstrating that the graphene is of good quality. A more detailed explanation of the graphene quality is provided in the Supplementary Information. The fabricated sensors are placed in ceramic packages and wire-bonded (Figure1c). Figure 1c further shows a color enhanced scanning electron micrograph (SEM) of a device, with the single-layer graphene patch on top of the $SiO_2$, the gold contacts underneath the graphene patch and the bond wires connected to the bond pads.

The experiments were carried out inside two separate chambers: A vacuum chamber operating (Figure 1d) at pressures below atmospheric pressure and a humidity chamber (Figure 1e) operating at atmospheric pressure. The two chambers were required to cover the full humidity range from 1% relative humidity (RH) to 96% RH. In the vacuum chamber, air is pumped out of the chamber, which reduces the water vapor (*i.e.* the humidity level) from about 30% RH down to 1% RH. The rate at which the pump vaccums air out of the chamber is controlled by a valve. The vacuum chamber was also used to expose the graphene devices to individual gasses in a controlled way. This allows evaluation of the cross-sensitivity of the sensor to competing gasses. In contrast, the humidity chamber operates at constant atmospheric pressure and can be filled with water vapor to vary the humidity from approximately 30% RH to 96% RH. This is done through the use of a humidifier which pumps water vapor into the chamber through a pipe. The water vapor flow rate from the humidifier is controlled by a dial on its side. A real time electrical readout in labview shows the current device resistance and chamber humidity at all times during the measurement. While precise control of the chamber humidity is difficult, observation of the relative changes in humidity measured with respect to resistance provides sufficient precision. During all measurements, the humidity is continuously monitored using a commercial HIH-4000 humidity sensor (Honeywell International Inc.). In addition, the gas pressure in the vacuum chamber is continuously monitored using a commercial digital vacuum transducer PDR 900 (MKS Instruments). Further, the temperature is monitored using a LM35 commercial temperature sensor. In experiments where temperature is monitored, graphene devices show little or no temperature dependency – neither from convective cooling from incoming air flow nor from Joule heating (See Supplementary Information).

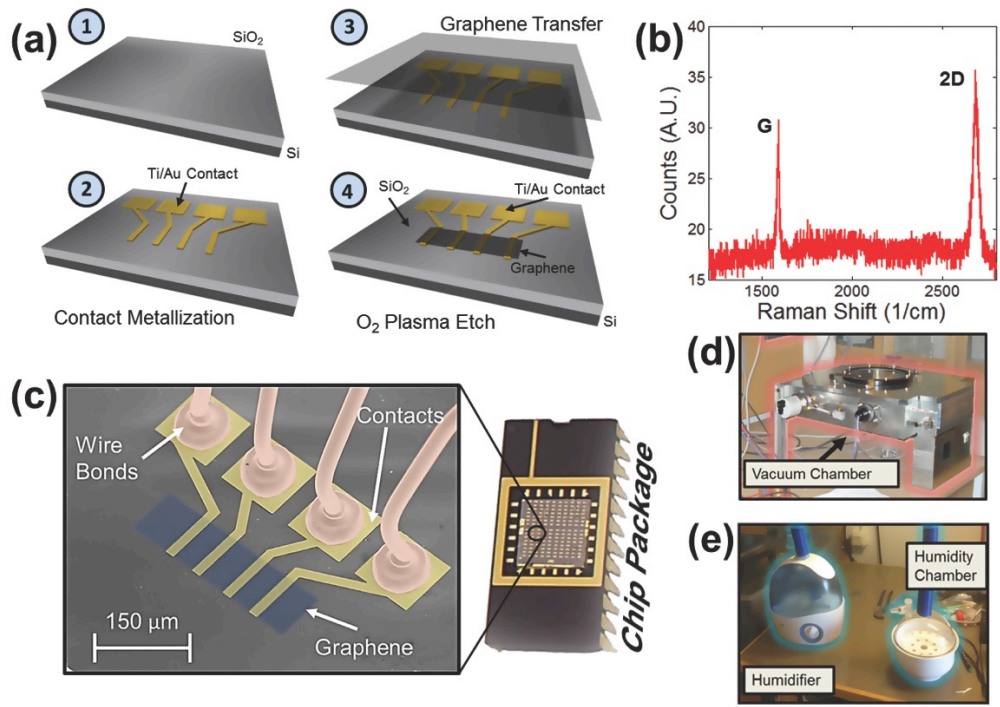

**Figure 1:** a) Process flow of the humidity sensor fabrication. (1) depicts a silicon substrate with 300 nm of SiO$_2$ thermally grown on the top surface. Cavities were etched and filled with 20 nm of Ti and 200 nm of Au in order to define electrical contacts to the graphene patch (2). Graphene was then transferred to the chip (3) and patterned using a photoresist mask and O$_2$ plasma etching (4). b) Raman spectrum of the graphene with distinctive G and 2D peaks. c) Color enhanced SEM image of a wire bonded device where the graphene, the contacts and the wire bonds are shaded in light blue, gold and orange, respectively. The packaged and wire-bonded devices inside a ceramic package are shown on the right. d) Vacuum chamber setup. e) Humidity chamber setup.

Figure 2a shows the measured relative change in resistance of the graphene sensor as the pressure chamber is evacuated (region 1, black symbols). The data is combined with the measured relative change in resistance of the graphene sensor in the separate humidity chamber as the humidity is increased (region 2, blue symbols). Note that, once the humidity begins to change, the resistance changes correspondingly and the response to change in humidity shows similar behavior in both chambers. As the humidity is decreased in the vacuum chamber by reducing the gas pressure, the corresponding resistance increases (Figure 2a-1). Likewise, as the humidity is increased in the humidity chamber by introducing water vapor, the corresponding resistance decreases (Figure 2a-2). Figure 2b shows a schematic of the interaction of water vapor with the graphene. As the humidity increases, more water molecules are adsorbed on the surface. Likewise, a decrease in humidity will cause water molecules to be desorbed from the surface. For measuring the resistance of the graphene patch, the device is placed in a Wheatstone bridge configuration. The graphene patches are biased with square-wave pulses of 200 mV and a pulse duration of 500 µs. This is done in order to mitigate drift in the device due to excessive Joule heating. Figure 2c compares measurement data from a graphene sensor placed in the vacuum chamber (black line) with the measured relative humidity shown on the figure as a %RH value using the commercial humidity sensor (red line) and the measured chamber pressure using the commercial pressure sensor (blue line). The output signals of the graphene humidity sensor, the commercial humidity senor and the pressure are remarkably similar - confirming that the changes in chamber pressure correlate with changes in water vapor concentration in the air.

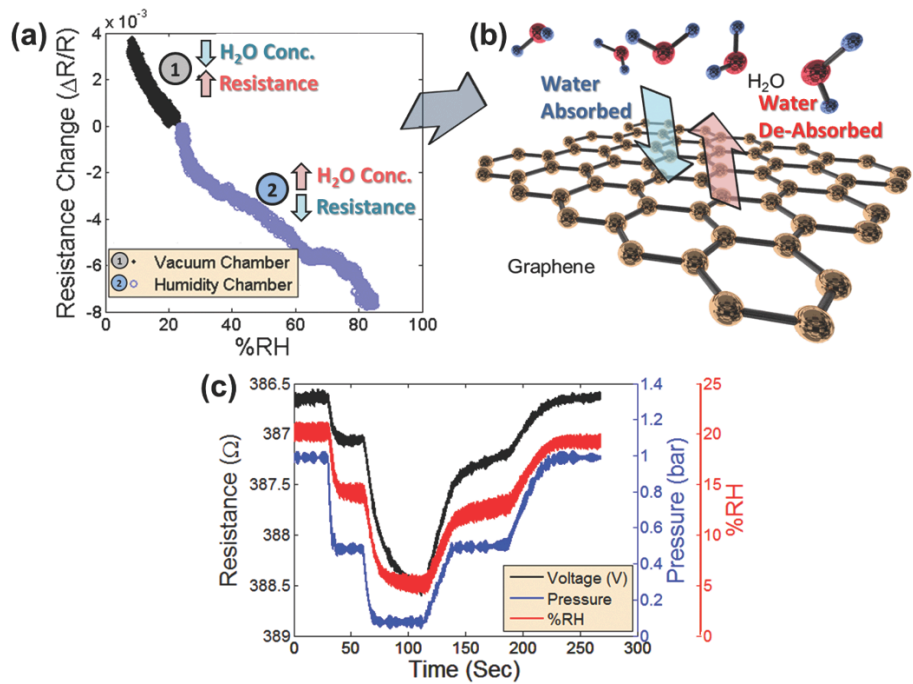

Figure 2: a) Resistance change in the graphene device versus the relative humidity (%RH) for a device placed in the vacuum chamber (1) and the same device placed in the humidity chamber (2). b) Interaction of water molecules with graphene surface. c) Resistance response (black lines) of the graphene device in a Wheatstone bridge configuration in comparison with the %RH response from both a commercial HIH-4000 humidity sensor (red line) as well as pressure response (bar) from a commercial PDR 900 pressure sensor (blue line) placed in the vacuum chamber.

A control experiment was conducted to rule out cross-sensitivities to pressure variations or to gasses typically present in air. Therefore, graphene devices were individually exposed to the most common gasses comprising air, including dry nitrogen ($N_2$), dry oxygen ($O_2$), and dry argon (Ar), as well as humid air. To achieve this, the pressure valve which controls the influx of air into the vacuum chamber was connected to a tank containing these pure and dry compressed gases. Once connected, the valve to the gas tank was opened and the gas was allowed to enter the chamber. By evacuating the pressure chamber to a vacuum of less than 200 mbar for several minutes while allowing the influx of pure dry gas, the chamber was filled with the gas. For each gas, the resistance change in the graphene sensor was then recorded while varying the chamber pressure between 200 mbar and 1 bar. The pressure was controlled through a valve between the chamber and the vacuum pump (similar to the experiments reported in Smith *et al.*).[32] The response of the graphene sensor as a result of varying pressure for each individual gas was such determined and is summarized in Figure 3a. Note that there is no significant change in resistance for any of the gasses tested individually. In contrast, the resistance change of graphene devices exposed to humid air persists among repeated trials. This strongly suggests that there is no influence of the individual gasses on the resistance of the graphene, but that it is indeed the humidity that is sensed.

The insensitivity of the sensor to the main gasses constituting air was further verified by connecting multiple gas supplies (dry Ar, dry $N_2$, dry $O_2$ and humid air) to the vacuum chamber and allowing each of the gasses to subsequently enter into the chamber. This confirms that none of the individual dry gasses has an effect on the graphene device when first introduced into the chamber. Figure 3b shows the resistance evolution of the graphene sensor (black dots) measured as each gas was introduced into the chamber. First, the chamber was filled with Ar and then Ar was circulated through the chamber for about 60s while the resistance response of the sensor was recorded. Then the argon flow was switched off and the chamber was evacuated. This procedure was repeated for $N_2$ and $O_2$. Finally, humid air was introduced in the chamber. Figure 3b shows no significant resistance response of the graphene sensor while each gas is introduced into the chamber, except for the case of humid air. The corresponding %RH response of the commercial HIH-4000 humidity sensor (red dots) in Figure 3b confirms that the humidity in the chamber is not significantly affected as dry gasses are being carefully circulated. This demonstrates that, within the time-scales of this study, each individual gas has very little effect on the resistance of the graphene device and that the graphene sensor has high-specificity for humidity in relation to the main gas constituents in air. This is in contrast to previous studies, which have report sensitivity of graphene to $O_2$, suggesting that there may be possible cross sensitivity[33]. However, that study was carried out over larger timescales with slower response times than reported here, which may serve to explain the different conclusions."

The situation is slightly different when there is a substantial gas flow across the sensor, as there may be competing events. First, an $N_2$ gun was used to control the flow of nitrogen over the surface of the sensor. As the $N_2$ flows over the surface of the device, the resistance increases noticeably and reproducibly (Figure 3c). This can be explained by a reduction of the surface concentration of water molecules on the graphene as illustrated in Figure 3d. In a similar fashion, the device was exposed to a flow of exhaled breath with a relative humidity of approximately 100%. Figure 3e shows the resistance response of a device while inhaling and exhaling breath several times in close proximity (within 10 cm) of the device. Here, exhaling (= flow on) leads to a decrease of the resistance. This result is in line with our expectations, because the device response is triggered by water vapor contained in the breath as illustrated in Figure 3f. In both cases, the exposure to $N_2$ flow and the exposure to breath, an increase in water vapor concentration causes a decrease in the resistance of the graphene device, while a decrease in the concentration of water vapor causes an increase in resistance of the graphene device. These results are expected and consistent with the results from the measurements in the vacuum and in the humidity chambers shown in Figure 2.

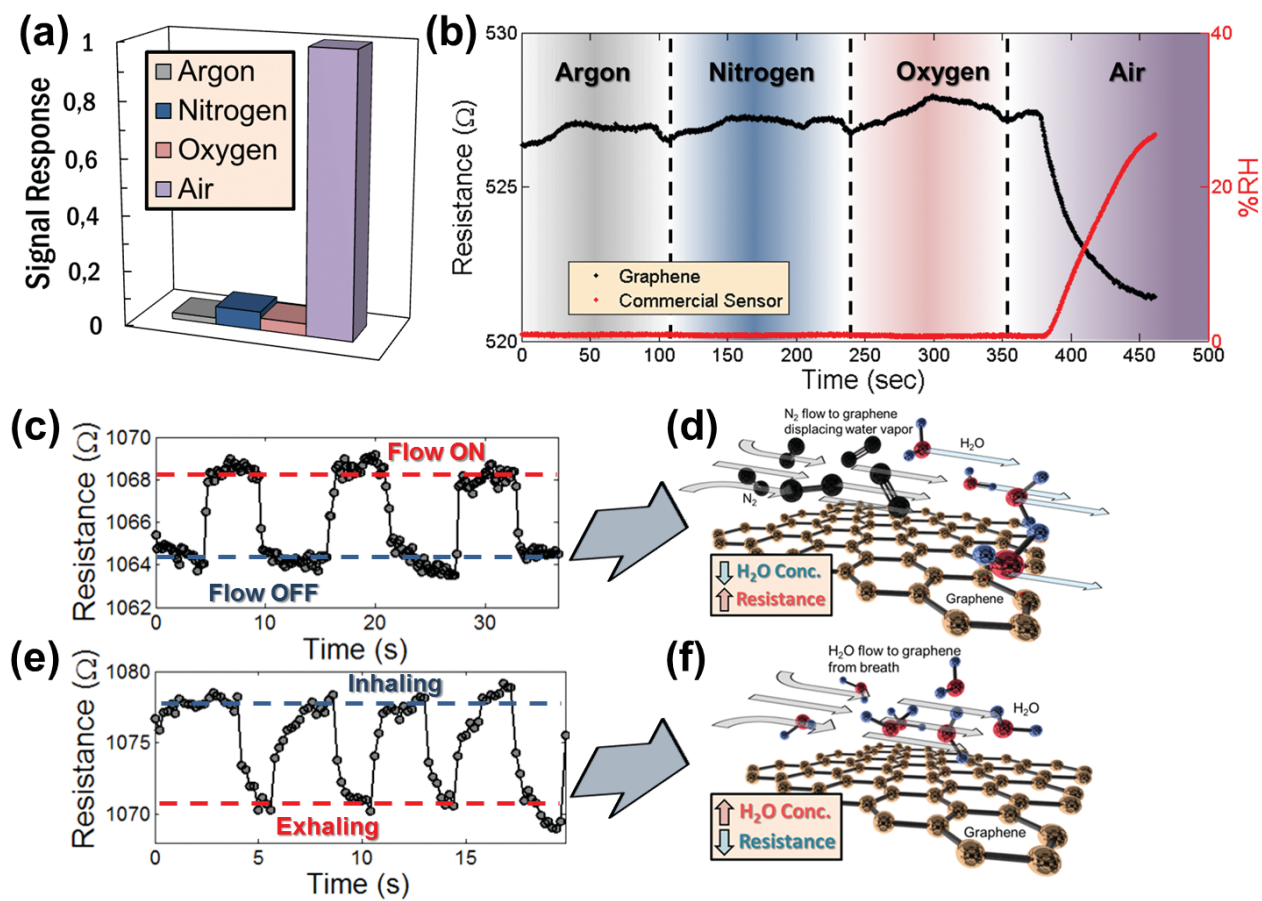

Figure 3: a) Normalized signal response of the graphene device for different gasses at varying pressures (resulting in varying humidity in the case of air). Note that the resistance change in the graphene devices is very small for any of the gasses, except for the case of air, which is the only gas containing an appreciable amount of water vapor. b) Change of resistance is a graphene device exposed subsequently to various gasses (black dots) compared with the relative humidity levels (%RH) measured by a commercial HIH-4000 sensor (red dots). Note that individual dry gasses have no effect on the resistance response of the graphene device. c) Graphene device in air and subject to pulses of $N_2$ flow. c) Schematic representation of how $N_2$ flow over the device affects the local humidity levels. e) Resistance response of a graphene device when exposed to inhaling and exhaling human breath. f) Schematic representation of water vapor being blown onto the device as it is breathed on, thereby increasing the local humidity levels.

The repeatability is addressed in Figure 4a, which shows the average measured graphene device resistance plotted versus relative humidity (%RH (red dots with a line). The error bars indicate the standard deviation of the data from the averaged value, with a maximum standard deviation of 0.1082 Ω. The gray dots represent the raw data. This data represents three cycles of pumping and venting of humid air in and out of the pressure chamber. Figure 4b compares the device conductivity, calculated with a simple resistor model based on the measured device dimensions and the known thickness of single-layer graphene of approximately 3.4 Å,[34] over time with the signal from the commercial humidity sensor. The data represents three cycles of modulating the humidity inside the vacuum chamber and shows remarkably consistent readout. Previously reported data for humidity sensing using graphene suggests very slow response and recovery times on the order of 180 s.[17] Figure 4c and 4d show the time resolved response and recovery of the graphene sensor and the commercial humidity sensor. The response and recovery times are defined here as the time it takes for the signal to reach from 10% of the initial humidity value to 90% of the final humidity value. In Figure 4c and 4d, the black dots represent the signal from the graphene sensor and the red dots represent the signal from the commercial humidity sensor. The 10% to 90% regions are marked with dashed black and red lines. Because the pumping and venting of water vapor in and out of the vacuum chamber involves a delay on the order of several seconds, it is not possible from measurements performed in the vacuum chamber to precisely determine the absolute response and recovery times using this experimental set-up. Therefore, an $N_2$ gun was used to induce a more local and rapid change of surface water molecules (compare Figure 3c and Figure 3d). Figure 4e and 4f show the response and recovery times in a graphene sensor from a flow-ON state to a flow-OFF state and vice versa. The 10% to 90% regions are again marked with dashed red and blue lines. The measured response times are 600 ms to 800 ms and the recovery times are 400 ms to 1 s. Figure 4g displays the resistance measurement of a device in the vacuum chamber from Figure 4b in high time resolution, just when the chamber humidity begins to increase. The graphene sensor (black dots) clearly responds faster than the commercial humidity sensor (red circles). The shaded black and red regions represent the approximate time at which a shift from decreasing to increasing humidity is measured. Here, the graphene device responds approximately 1-2 seconds faster than the commercial sensor. The combination of these measurements suggest that the response and recovery times of the devices are on the order of milliseconds and are possibly much faster than observed due to limitations in the experimental setup.

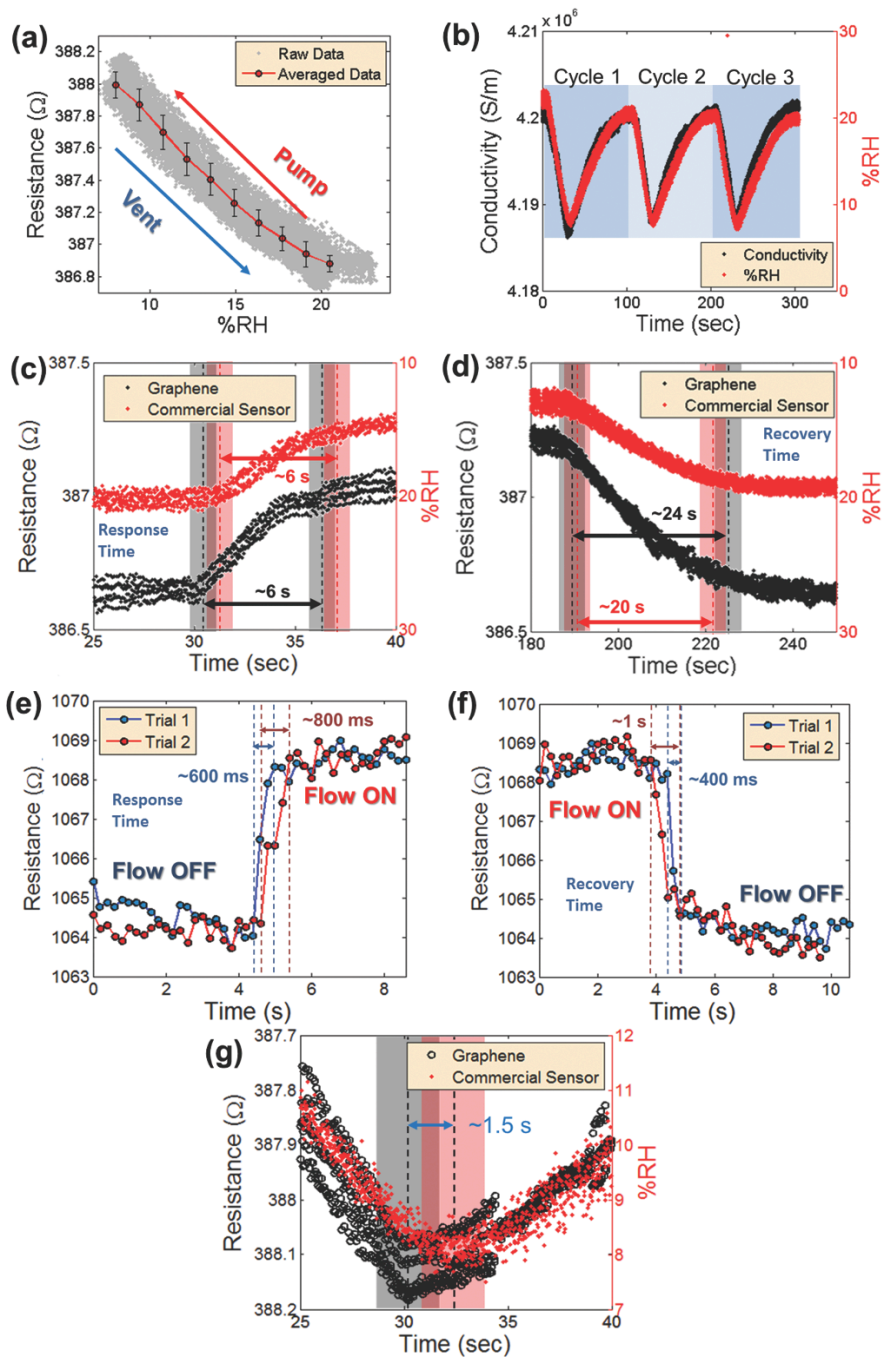

**Figure 4:** a) Resistance versus relative humidity (%RH) of a graphene device during pumping and venting of the vacuum chamber. The inset shows the corresponding conductivity of the same device. b) Conductivity of a graphene device measured in conjunction with measuring the %RH versus time, using the commercial humidity sensor. The data represents 3 measurement cycles where the humidity is varied by evacuating and venting the vacuum chamber with air. c) Resistance response versus time of a graphene device compared with the **commercial humidity sensor to characterize the response time of the device. d) Resistance response versus time of a device compared with the commercial humidity sensor to evaluate the recovery time of the device. e) Response time of a graphene device when gas flow is introduced using a $N_2$ gun. f) Corresponding recovery time when the device is subjected to a gas flow. g) Close-up of the resistance output of the graphene device compared with the commercial humidity sensor. Note that the graphene device responds 1-2 s faster than the commercial humidity sensor.**

The mechanism underpinning graphene's sensitivity to humidity may be the result of an electrostatic interaction between the water and the graphene. Of all the gasses comprising air, water is the only one containing a dipole. Thus, simulation of the water/graphene interaction can provide insight into the sensitivity mechanism. A number of studies have previously investigated the effect of the presence of water molecules on the surface of graphene on $SiO_2$ substrates.[35–39] Building upon these investigations, ground-state density functional theory calculations for graphene in different configurations of humid environments were performed. The graphene-water system is modelled as single water molecules that are arranged in a monolayer of water above a single-layer graphene sheet with a separation of 3.5 Å. The graphene sheet rests above a layer of $SiO_2$. The experimentally observed change in conductivity in the graphene suggests that the water molecules dope the graphene layer. We therefore performed density functional calculations on the graphene-water system for different cases: a perfect $SiO_2$ substrate and an $SiO_2$ substrate with a well-established surface defect, a $Q_3^0$ defect.[40,41] The distance between the $Q_3^0$ silicon atom and the graphene sheet is set to 4.1 Å in our simulations. The surface defect was then incorporated into the simulation for the two relevant cases, *i.e.* with and without the monolayer of water molecules. The unit cell of the simulation is shown in Figure 5a. The red dotted line denotes the cutting plane through which the charge density difference is examined. The $Q_3^0$ defect is introduced into the system as illustrated in Figure 5b. The figure shows a contour map of the charge density difference of the defected system without the presence of a water molecule. Likewise, Figure 5c shows a contour map of the charge density difference with the presence of the water molecule and the $Q_3^0$ defect. When a defect is present in the $SiO_2$ surface, a substantial charge transfer and dipole moment is formed in the $SiO_2$ layer. Figure 5d shows the perfect system (*i.e.* no defect in the $SiO_2$ is present). For the case of the perfect system, a hydrogen atom is used to passivate the dangling oxygen atomic bond in the substrate. When both the $Q_3^0$ defect and the water molecules are present (Figure 5c), the charge density between the graphene layer and the water molecules becomes significantly different as compared to the system with $Q_3^0$ defects but without water (Figure 5b) as well as the perfect system (Figure 5d). In the calculations, the $SiO_2$ surface defects give rise to an impurity band, similar to the results of Wehling *et al.*[39] The electrostatic dipole moment of the water molecules may now shift this impurity band, leading to an effective doping and increased conductivity in the graphene layer, which is in line with the experimental observations. However, the simulations reveal that the humidity sensing effect is due to the graphene layer contacting both the defected $SiO_2$ layer while being influenced by water. Further, freestanding graphene may not be sensitive to water molecules[38]. This is an interesting prediction which suggests that graphene's substrate can effectively functionalize the material to become more sensitive to a specified adsorbent.

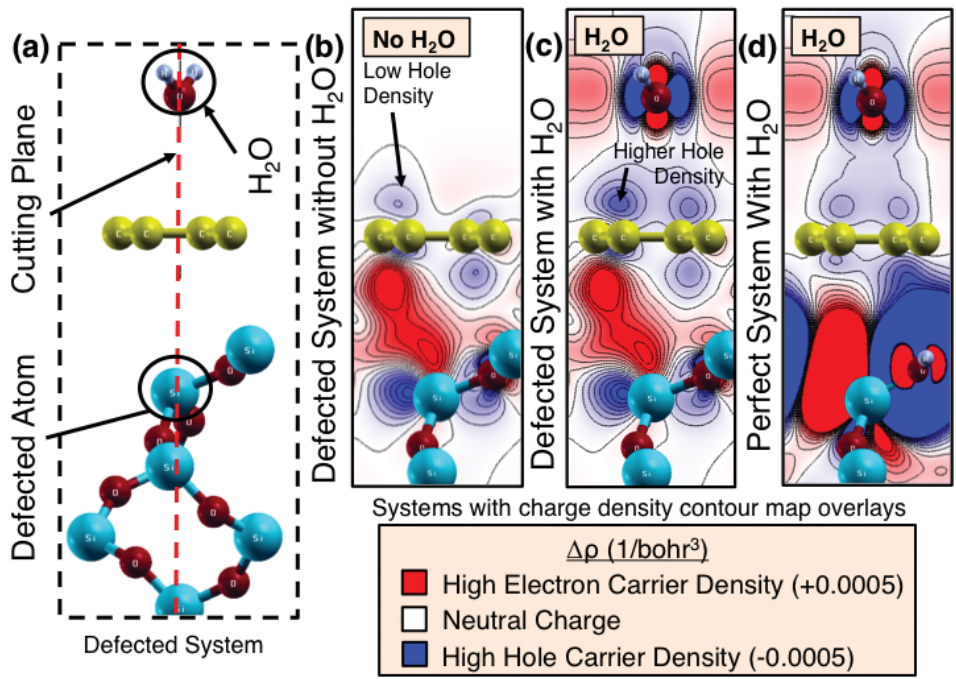

**Figure 5: Charge-density difference (CDD) plots for the three simulated systems.** The CDD is calculated by subtracting the charge density of the corresponding subsystems from the calculated charge density of the system. a) Unit cell of graphene on top of SiO2 with $Q_3^0$ defects, with water added on top of the graphene layer. b) CDD for graphene on top of SiO2 with $Q_3^0$ defects present. No water. c) CDD for the same system as in b), but with water added on top of the graphene layer. d) CDD for graphene on top of defect-free SiO2, with water added on top of the graphene layer. Note the charge accumulation at the graphene surface in panel c) where both $Q_3^0$ defects and water are present.

In order to benchmark the investigated resistive graphene humidity sensors against a commercial product and other suggested potential nanotechnologies, Figure 6a shows a comparison with the characterized humidity range of a number of different humidity sensors reported in literature.[4,14–16,42,43] The graphene sensor in the present study has been characterized for a larger humidity range than any other experimental device in literature. Figure 6b compares the response times of the presented graphene device with that of graphene oxide (GO) and tin oxide ($SnO_2$) resistive sensors. Note that while both GO and graphene outperform $SnO_2$, GO appears to be a superior sensor with respect to both response and recovery times. However, this originates rather in limitations in the measurement setup and the general difficulty in determining precise response and recovery times than the graphene sensor itself. Finally, the sensitivity of the graphene humidity sensor was calculated using Eq. 1,

$$S = \frac{\Delta R}{R \cdot \Delta \%RH} \cdot 100 \qquad (1).$$

Here, S is defined at the percent change in resistance divided by the percent change in relative humidity. Figure 6c compares data for different sensor technologies with the graphene sensors investigated in this paper, in particular for the relative humidity (%RH) range, response time, recovery time, and sensitivity.[4,13–18,42,43] It should be noted that the sensitivities reported for the different emerging humidity sensor technologies are not directly comparable and are therefore only indicative. For example, if the sensitivity of our sensor is compared directly as a simple change in resistance relative to the absolute resistance (without considering the humidity range measured), our values are then comparable those reported by Ghosh[17]. However, without a consistent measurement range, the sensitivities cannot be adequately compared and are therefore simply reported as measured. Further, the comparably small sensitivity of the graphene layer could be due to less defects in the graphene layer. GO in contrast has a large degree of dangling bonds which could contribute to its sensitivity, albeit at the expense of a higher resistance. For example, graphene based pressure sensors have a resistance on the order of 100s of Ohms to approximately 1kΩ. By comparison, the GO sensors have a resistance on the order of 1 to 10 MΩ[14]. This suggests graphene would be well suited for low power devices. Further, previous reports have explored how grain size or grain boundary density can affect the transport properties in graphene[44–47] with higher density leading to enhanced chemical sensing[45,46]. Thus, tuning the density of the grain boundaries may be considered to further improve the sensitivity. While GO-based sensors seem to be the most viable sensors for resistive, capacitive, and piezoresistive sensing of humidity, the graphene-based resistive sensors investigated in this work perform well in comparison to other resistive humidity sensors. However, one further advantage of graphene over GO may be a greater degree of controllability during fabrication. GO is prepared chemically and there is therefore the possibility of the occurance of chemical impurities during its preparation.

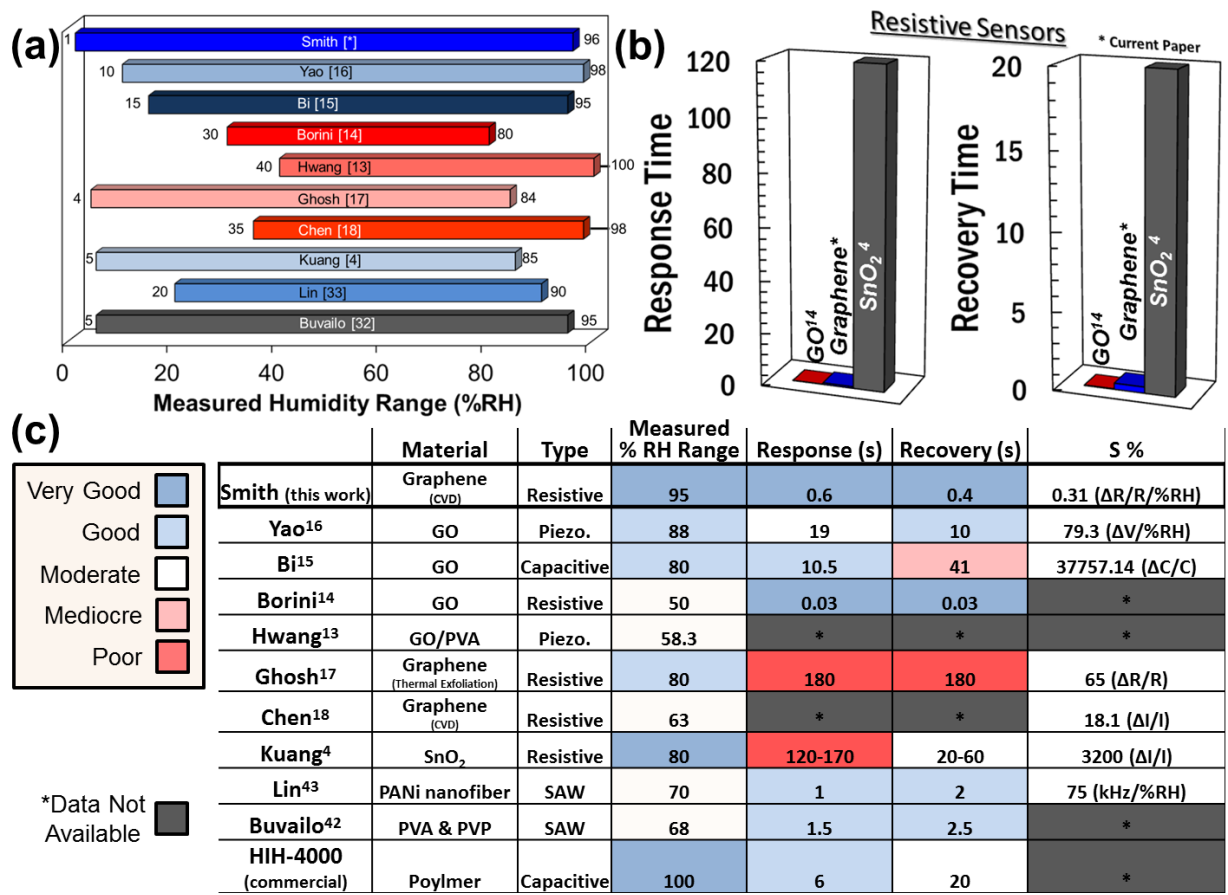

Figure 6: a) Humidity sensors reported in literature with the characterized humidity ranges. b) Response and recovery times for the presented graphene device compared with resistive humidity sensors reported in literature. c) Comparison of various humidity sensor technologies with respect to reported %RH ranges, response and recovery times, and sensitivities. Note that the HIH-4000 is a commercial sensor used for comparison in all experiments.

In conclusion, we have demonstrated and characterized resistive graphene humidity sensors with high specificity to other gas constituents present in air. The simulations performed in this work suggest that the sensitivity of the resistance of a graphene patch to water vapor results from the interaction between the water electrostatic dipole moment with impurity bands in the substrate. This effect in turn leads to doping of the graphene layer, causing increased conductivity as a result of the increased doping. We therefore propose that electrostatic dipole moments plays a key role in graphene doping and the related sensing mechanism, in particular since all other molecules studied here were free of electrostatic dipole moments. The graphene sensors show wide range sensitivity and good response and recovery times with values ranging below one second – competing with graphene oxide sensors and significantly outperforming previous graphene based humidity sensors. The simplicity of the device design using CVD graphene potentially offers a low cost, scalable technology that is integrable with back-end-of-the-line commercial semiconductor technology.

**Methods**

Devices consisting of graphene patches were fabricated on p-doped silicon substrates with a 300 nm thick $SiO_2$ layer. After thermal oxidation to 300 nm of $SiO_2$, four contact pads were embedded into the substrate. First, 200 nm deep cavities were etched into the $SiO_2$ layer with reactive ion etching (RIE) using Ar and $CHF_3$ gas at 200 W and 40 mTorr. Afterwards, 20 nm of titanium were evaporated to act as an adhesion layer, followed by the deposition of 200 nm of gold. The contacts were patterned using a lift-off process with the same self-aligned mask for RIE and metal deposition. The resulting contacts extended 20 nm above the substrate. This procedure limits the number of processing steps after the graphene transfer, as particularly the lift-off process can damage the graphene. The bond pads on the chips are 100 μm by 100 μm in order to allow sufficient area for the wire bonds. At this point, the wafers were diced into chip size. A layer of graphene was then deposited on the surface of the chip using a CVD graphene wet transfer technique.[48] An added advantage of depositing the graphene in the final process step before device packaging and characterization, is that the risk for damaging and contaminating the graphene by subsequent process steps is reduced. The CVD graphene we used was both grown in house as well as commercially available graphene on copper foils. The graphene in both cases is of comparable quality and a more detailed analysis of the graphene quality is given in the Supplementary Information. A layer of poly(Bisphenol A) carbonate (PC) was spin-coated onto the front side of the foil in order to act as a carrier layer.[49–53] Carbon residues on the backside of the copper foil were removed using an $O_2$ plasma etch. The copper foil was then etched in ferric chloride ($FeCl_3$), leaving the graphene / polymer layer floating in the solution.

The graphene was then cleaned of Fe ions using 8% HCl by volume and water. Next, the graphene was transferred from the solution to the chip. The chip was subsequently dried on a hot plate at 45 °C for 5 minutes. It was then left in chloroform for about 12 hours in order to remove the polymer carrier layer. The graphene was then patterned using a photoresist mask (SPR 700 1.2) and $O_2$ plasma for etching of the exposed graphene. After removal of the photoresist in acetone / isopropanol, the chips were wire bonded into chip packages (Figure 1c) to allow for reliable experiments in various chambers and environments. All experiments are performed at room temperature.

Simulations have been performed using ground-state density functional theory calculations for graphene existing in different configurations of humid environments. The model is comprised of single water molecules arranged in a monolayer on top of the graphene sheet, with 3.5Å as the distance between the graphene and the monolayer of water molecules. The Kohn-Sham equations have been solved using a plane wave basis set[54] with cut-off at 130 Ry and norm-conserving pseudo-potentials as implemented in the Quantum Espresso (QE) simulation package.[55] The Perdew, Burke and Ernzerhof (PBE) functional[56] approximation to the exchange-correlation part of the density functional was used. Hamann, Schluter, Chiang and Vanderbilt[57] (HSCV) norm-conserving pseudo-potentials were used for all the atoms. Calculations of total energy self-consistent field (SCF) were sampled using a 16x16x1 k-point mesh.

**Acknowledgements**

Support from the European Commission through two ERC Starting Grants (InteGraDe, 307311 & M&M's, 277879), the Swedish Research Council (E0616001 and D0575901, 24 iGRAPHENE) as well as the German Research Foundation (DFG, LE 2440/1-1) is gratefully acknowledged. We further acknowledge financial support from Erasmus Target II, the Göran Gustafsson Foundation, VR (Vetenskapsrådet), KVA (The Royal Swedish Academy of Sciences), KAW (the Knut and Alice Wallenberg Foundation), CTS (Carl Tryggers Stiftelse), STEM (Swedish Energy Agency), and SSF (Swedish Foundation for Strategic Research). The computations were performed on resources provided by the Swedish National Infrastructure for Computing (SNIC) at the PDC center for high-performance computing, KTH.

**Additional Information**

The authors declare no competing financial interests.

**Author Contribution Statement**

A.D.S., F.N. and M.C.L. conceived the experiments. A.D.S. performed the experiments and drafted the manuscript. K.E. carried out the simulations, with support from A.D., L.B. and M.R., H.H., A.D. and K.E. analyzed the output from the simulations and wrote the theory part of the paper. A.C.F. contributed to optimization of the experimental setup. F.F. designed the pulse measurement setup used in the experiments. S.V. contributed device fabrication technology. S.S. performed wire bonding of all devices. S.K. grew the CVD graphene films. A.D.S, F.N., M.Ö. and M.C.L analyzed the experimental results and all authors co-wrote and reviewed the manuscript.

**References**

1. Moseley, P. T. Solid state gas sensors. *Meas. Sci. Technol.* **8,** 223–237 (1997).

2. Capone, S. *et al.* Solid state gas sensors: state of the art and future activities. *J. Optoelectron. Adv. Mater.* **5,** 1335–1348 (2003).

3. Korotcenkov, G. Metal oxides for solid-state gas sensors: What determines our choice? *Mater. Sci. Eng. B* **139,** 1–23 (2007).

4. Kuang, Q., Lao, C., Wang, Z. L., Xie, Z. & Zheng, L. High-sensitivity humidity sensor based on a single $SnO_2$ nanowire. *J. Am. Chem. Soc.* **129,** 6070–6071 (2007).

5. Kong, J. *et al.* Nanotube molecular wires as chemical sensors. *Science* **287,** 622–625 (2000).

6. Zhou, X. T. *et al.* Silicon nanowires as chemical sensors. *Chem. Phys. Lett.* **369,** 220–224 (2003).

7. Wang, Y., Jiang, X. & Xia, Y. A solution-phase, precursor route to polycrystalline $SnO_2$ nanowires that can be used for gas sensing under ambient conditions. *J. Am. Chem. Soc.* **125,** 16176–16177 (2003).

8. Basu, S. & Bhattacharyya, P. Recent developments on graphene and graphene oxide based solid state gas sensors. *Sens. Actuators B Chem.* **173,** 1–21 (2012).

9. Ratinac, K. R., Yang, W., Ringer, S. P. & Braet, F. Toward Ubiquitous Environmental Gas Sensors Capitalizing on the Promise of Graphene. *Environ. Sci. Technol.* **44,** 1167–1176 (2010).

10. Yuan, W. & Shi, G. Graphene-based gas sensors. *J. Mater. Chem. A* **1,** 10078–10091 (2013).

11. Zhou, Y., Jiang, Y., Xie, T., Tai, H. & Xie, G. A novel sensing mechanism for resistive gas sensors based on layered reduced graphene oxide thin films at room temperature. *Sens. Actuators B Chem.* **203,** 135–142 (2014).

12. Dan, Y., Lu, Y., Kybert, N. J., Luo, Z. & Johnson, A. C. Intrinsic response of graphene vapor sensors. *Nano Lett.* **9,** 1472–1475 (2009).

13. Hwang, S.-H., Kang, D., Ruoff, R. S., Shin, H. S. & Park, Y.-B. Polyvinyl Alcohol Reinforced and Toughened with Poly (dopamine)-Treated Graphene Oxide, and Its Use for Humidity Sensing. *ACS Nano* **8,** 6739–6747 (2014).

14. Borini, S. *et al.* Ultrafast graphene oxide humidity sensors. *ACS Nano* **7,** 11166–11173 (2013).

15. Bi, H. *et al.* Ultrahigh humidity sensitivity of graphene oxide. *Sci. Rep.* **3,** (2013).

16. Yao, Y., Chen, X., Guo, H., Wu, Z. & Li, X. Humidity sensing behaviors of graphene oxide-silicon bi-layer flexible structure. *Sens. Actuators B Chem.* **161,** 1053–1058 (2012).

17. Ghosh, A., Late, D. J., Panchakarla, L. S., Govindaraj, A. & Rao, C. N. R. $NO_2$ and humidity sensing characteristics of few-layer graphenes. *J. Exp. Nanosci.* **4,** 313–322 (2009).

18. Chen, M.-C., Hsu, C.-L. & Hsueh, T.-J. Fabrication of Humidity Sensor Based on Bilayer Graphene. *Electron Device Lett.* **35,** 590–592 (2014).

19. Kim, H.-Y., Lee, K., McEvoy, N., Yim, C. & Duesberg, G. S. Chemically modulated graphene diodes. *Nano Lett.* **13,** 2182–2188 (2013).

20. Bolotin, K. I. *et al.* Ultrahigh electron mobility in suspended graphene. *Solid State Commun.* **146,** 351–355 (2008).

21. Morozov, S. V. *et al.* Giant intrinsic carrier mobilities in graphene and its bilayer. *Phys. Rev. Lett.* **100,** 016602 (2008).

22. Massera, E. *et al.* Gas sensors based on graphene. *Chem. Today* **29,** 39–41 (2011).

23. Fowler, J. D. *et al.* Practical chemical sensors from chemically derived graphene. *ACS Nano* **3,** 301–306 (2009).

24. Gautam, M. & Jayatissa, A. H. Graphene based field effect transistor for the detection of ammonia. *J. Appl. Phys.* **112,** 064304 (2012).

25. He, Q., Wu, S., Yin, Z. & Zhang, H. Graphene-based electronic sensors. *Chem. Sci.* **3,** 1764–1772 (2012).

26. Wehling, T. O. *et al.* Molecular doping of graphene. *Nano Lett.* **8,** 173–177 (2008).

27. Schedin, F. *et al.* Detection of individual gas molecules adsorbed on graphene. *Nat. Mater.* **6,** 652–655 (2007).

28. Yang, Y., Brenner, K. & Murali, R. The influence of atmosphere on electrical transport in graphene. *Carbon* **50,** 1727–1733 (2012).

29. Moser, J., Verdaguer, A., Jiménez, D., Barreiro, A. & Bachtold, A. The environment of graphene probed by electrostatic force microscopy. *Appl. Phys. Lett.* **92,** 123507 (2008).

30. Kataria, S. *et al.* Chemical vapor deposited graphene: From synthesis to applications (Phys. Status Solidi A 11∕2014). *Phys. Status Solidi A* **211,** n/a–n/a (2014).

31. Ferrari, A. C. *et al.* Raman spectrum of graphene and graphene layers. *Phys. Rev. Lett.* **97,** 187401 (2006).

32. Smith, A. D. *et al.* Electromechanical piezoresistive sensing in suspended graphene membranes. *Nano Lett.* **13,** 3237–3242 (2013).

33. Chen, C. W. *et al.* Oxygen sensors made by monolayer graphene under room temperature. *Appl. Phys. Lett.* **99,** 243502 (2011).

34. Gupta, S. S. Elastic constants from molecular mechanics simulations of frequencies of free-free single-walled carbon nanotubes and clamped single-layer graphene sheets. (2009).

35. Yang, Y. & Murali, R. Binding mechanisms of molecular oxygen and moisture to graphene. *Appl. Phys. Lett.* **98,** 093116 (2011).

36. Freitas, R. R. Q., Rivelino, R., Mota, F. & de Castilho, C. M. C. DFT Studies of the Interactions of a Graphene Layer with Small Water Aggregates. *J. Phys. Chem. A* **115,** 12348–12356 (2011).

37. Leenaerts, O., Partoens, B. & Peeters, F. M. Water on graphene: Hydrophobicity and dipole moment using density functional theory. *Phys. Rev. B* **79,** 235440 (2009).

38. Wehling, T. O., Katsnelson, M. I. & Lichtenstein, A. I. Adsorbates on graphene: Impurity states and electron scattering. *Chem. Phys. Lett.* **476,** 125–134 (2009).

39. Wehling, T. O., Lichtenstein, A. I. & Katsnelson, M. I. First-principles studies of water adsorption on graphene: The role of the substrate. *Appl. Phys. Lett.* **93,** 202110–202110–3 (2008).

40. Wilson, M. & Walsh, T. R. Hydrolysis of the amorphous silica surface. I. Structure and dynamics of the dry surface. *J. Chem. Phys.* **113,** 9180–9190 (2000).

41. Walsh, T. R., Wilson, M. & Sutton, A. P. Hydrolysis of the amorphous silica surface. II. Calculation of activation barriers and mechanisms. *J. Chem. Phys.* **113,** 9191–9201 (2000).

42. Buvailo, A., Xing, Y., Hines, J. & Borguet, E. Thin polymer film based rapid surface acoustic wave humidity sensors. *Sens. Actuators B Chem.* **156,** 444–449 (2011).

43. Lin, Q., Li, Y. & Yang, M. Highly sensitive and ultrafast response surface acoustic wave humidity sensor based on electrospun polyaniline/poly (vinyl butyral) nanofibers. *Anal. Chim. Acta* **748,** 73–80 (2012).

44. Roche, S. Fundamentals of Charge Transport in Polycrystalline graphene and hybrids. (2015).

45. Yasaei, P. *et al.* Chemical sensing with switchable transport channels in graphene grain boundaries. *Nat. Commun.* **5,** (2014).

46. Seifert, M. *et al.* Role of grain boundaries in tailoring electronic properties of polycrystalline graphene by chemical functionalization. *2D Mater.* **2,** 024008 (2015).

47. Duong, D. L. *et al.* Probing graphene grain boundaries with optical microscopy. *Nature* **490,** 235–239 (2012).

48. Lapisa, M., Stemme, G. & Niklaus, F. Wafer-level heterogeneous integration for MOEMS, MEMS, and NEMS. *IEEE J. Sel. Top. Quantum Electron.* **17,** 629–644 (2011).

49. Reina, A. *et al.* Transferring and identification of single-and few-layer graphene on arbitrary substrates. *J. Phys. Chem. C* **112,** 17741–17744 (2008).

50. Li, X. *et al.* Transfer of large-area graphene films for high-performance transparent conductive electrodes. *Nano Lett.* **9,** 4359–4363 (2009).

51. Jiao, L. *et al.* Creation of nanostructures with poly (methyl methacrylate)-mediated nanotransfer printing. *J. Am. Chem. Soc.* **130,** 12612–12613 (2008).

52. Park, H. J., Meyer, J., Roth, S. & Skákalová, V. Growth and properties of few-layer graphene prepared by chemical vapor deposition. *Carbon* **48,** 1088–1094 (2010).

53. Lin, Y.-C. *et al.* Clean transfer of graphene for isolation and suspension. *ACS Nano* **5,** 2362–2368 (2011).

54. Pickett, W. E. Pseudopotential methods in condensed matter applications. *Comput. Phys. Rep.* **9,** 115–197 (1989).

55. Giannozzi, P. *et al.* QUANTUM ESPRESSO: a modular and open-source software project for quantum simulations of materials. *J. Phys. Condens. Matter* **21,** 395502 (2009).

56. Perdew, J. P., Burke, K. & Ernzerhof, M. Generalized gradient approximation made simple. *Phys. Rev. Lett.* **77,** 3865 (1996).

57. Vanderbilt, D. Optimally smooth norm-conserving pseudopotentials. *Phys. Rev. B* **32,** 8412 (1985).31